\documentclass[times,onecolumn]{elsarticle} 

\usepackage{array,booktabs} 
\newcommand {\otoprule}{\midrule [\heavyrulewidth]} 
\newcommand{\lightmidrule}{\midrule [0.2pt]}

\usepackage[a4paper,hscale=0.72,vscale=0.7,centering]{geometry}
\usepackage{amssymb,amsmath}%

\usepackage{graphics,times,xspace}

\usepackage[caption=false]{subfig}

\usepackage{hyperref}

\IfFileExists{srcltx.sty}{\usepackage[active]{srcltx}}

 \def\fermi{Fermi\xspace}%
\renewcommand{\deg}{\ensuremath{^\circ}\xspace}

\def\magenta{\textsc{Reg~1}\xspace} 
\def\second {\textsc{Reg~2}\xspace} %
\def\third {\textsc{Reg~3}\xspace}
\def\centerreg{\textsc{Central}\xspace} 

\begin{document}
\begin{frontmatter}
  \title{Spectral and spatial variations of the diffuse gamma-ray background
    in the vicinity of the Galactic plane and possible nature of the feature
    at 130
    GeV\\
    \rightline{\texttt{\small CERN-PH-TH-2012-141}}} \author[a,b] {Alexey
    Boyarsky} \author[b]{Denys Malyshev} \author[c]{Oleg Ruchayskiy}

\address[a]{Instituut-Lorentz for Theoretical Physics, Universiteit Leiden,
  Niels Bohrweg 2, Leiden, The Netherlands}

\address[b]{Bogolyubov Institute of Theoretical Physics, Kyiv, Ukraine} %

\address[c]{CERN Physics Department, Theory Division, CH-1211 Geneva 23,
  Switzerland}

\begin{abstract}
  We study the properties of the diffuse $\gamma$-ray background around the
  Galactic plane at energies 20 -- 200~GeV. We find that the spectrum of this
  emission possesses significant spacial variations with respect to the
  average smooth component.  The positions and shapes of these spectral
  features change with the direction on the sky.  We therefore argue, that the
  spectral feature around 130~GeV, found in several regions around the
  Galactic Center and in the Galactic plane in
  \cite{Brigmann:2012,weniger12,tempel12,Su-GC}, can not be interpreted with
  confidence as a $\gamma$-ray line, but may be a component of the diffuse
  background and can be of instrumental or astrophysical origin. Therefore,
  the dark matter origin of this spectral feature becomes dubious.
\end{abstract}

\end{frontmatter}

\section{Introduction}
It has been recently reported in~\cite{Brigmann:2012} and further investigated
in \cite{weniger12} and in~\cite{Su-GC}\footnote{The latter paper has appeared
  after the first version of the present work
  \href{http://arxiv.org/abs/arXiv:1205.4700v1}{1205.4700v1}} that the
$\gamma$-ray emission from the region around the Galactic Center (GC) exhibits
a line-like excess at the energies $\sim 130$~GeV.  An interest to this result
is based on the expectation that any signal of astrophysical origin at high
energies would have a broad (compared to the \fermi\ spectral resolution)
spectral shape. Diffuse emission with the line-like spectrum has therefore
been considered as an exotic one, e.g. as a ``smoking gun'' for dark matter
annihilation \cite{Bergstrom:1988fp} (see e.g.\
\cite{Bergstrom:2012fi,Bringmann:2012uy} for review). In particular, ``Higgs
in space'' scenario~\cite{Jackson:2009kg} predicts a $\gamma$-ray line at
$\sim 130$~GeV for the Higgs mass around $125$~GeV as seen by the
LHC~\cite{ATLAS,CMS}.

The region of~\cite{Brigmann:2012} and~\cite{weniger12} was selected by
maximizing signal-to-noise ratio for the expected dark matter annihilation
signal.  The preprints of~\cite{Brigmann:2012} and~\cite{weniger12} were
followed by~\cite{tempel12} where the claim was confirmed and it was
demonstrated that a similar excess originates from several regions of the size
$\sim 3$\deg around the Galactic plane.  A number of
works~\cite{tempel12,Profumo:2012tr,Dudas:2012pb,Ibarra:2012dw,Cline:2012,Aharonian:2012cs},
have discussed possible interpretations and origin of this spectral feature, see \cite{Bringmann:2012uy} for the review.
The search for $\gamma$-ray lines, based on the 2-year data, performed by the
\fermi\ collaboration~\cite{Ackermann:2012qk} did not reveal any lines but had
not comment on the origin of the observed excess.

It was demonstrated in the first version of this paper
(\href{http://arxiv.org/abs/1205.4700v1}{1205.4700v1}) that spectral features
with the significance, similar to that of the excess observed in the Galactic
center region around $130$~GeV can be also found at other energies in
different regions of the sky.  It was then argued
in~\cite{Hektor:12c,Finkbeiner:12a}\footnote{see also talks by C.~Weniger at
  IDM-2012~\cite{IDM} and COSMO-2012~\cite{COSMO} conferences} that the only
``significant feature'' is the one in the GC, while all the other features,
found in \href{http://arxiv.org/abs/1205.4700v1}{1205.4700v1} can be
considered as pure statistical fluctuations.

We believe that the current data do not allow to reach a definitive conclusion
here. Moreover, the most relevant question is actually different. To support
the DM interpretation of the GC line, one should try to determine:
\begin{itemize}[--]
\item Is it possible to \emph{exclude} the existence of features in several
  regions around the Galactic plane (apart from the GC)?
\end{itemize}
We demonstrate in this work that the answer to this question is
\emph{negative} -- the data does not allow to separate with confidence a
feature in the Galactic Center from other features in the Galactic
plane. Moreover, common interpretation of all the features is possible,
providing additional evidence for such a hypothesis.  Therefore that the
\emph{nature of all these deviations} from simple featureless spectral models
at energies 50--200 GeV should be discussed, rather then some peculiar
properties of the Galactic center region. Until this question is settled the
DM origin of the 130~GeV line (or a pair of lines at 110 and 130 GeV) remains
dubious.

\begin{figure*}[t]
  \includegraphics[width=\textwidth]{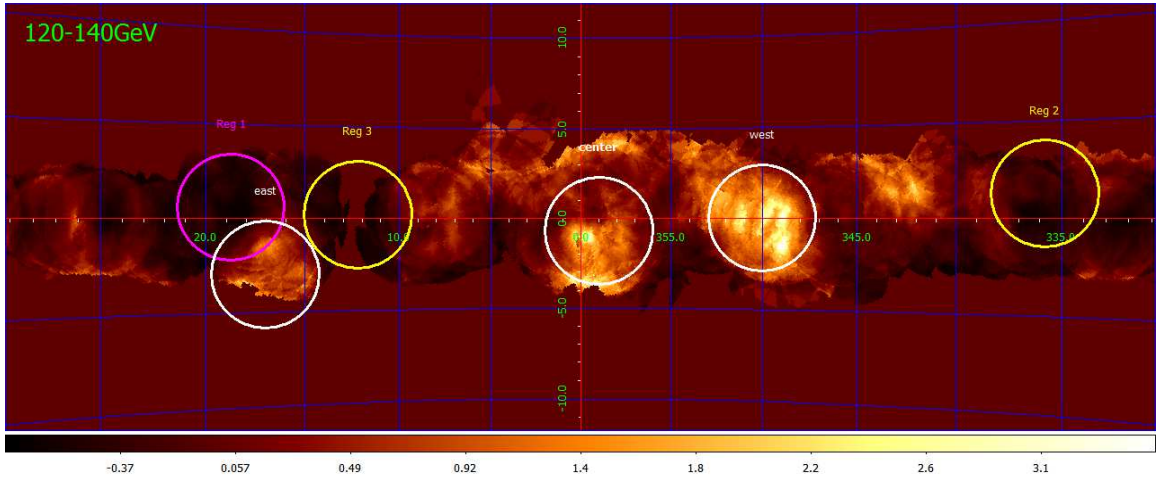}
  \caption{Difference in the number of photons between the energy bin
    120--140~GeV and the half-sum of two adjacent bins (100--120 and
    140--160~GeV). The regions with positive and negative excesses around the
    background are clearly visible. Three most significant regions from
    \cite{tempel12} are shown with white
    circles.  }
\label{fig:120_140GeV}
\end{figure*}

\begin{figure*}[t]
\includegraphics[width=\textwidth]{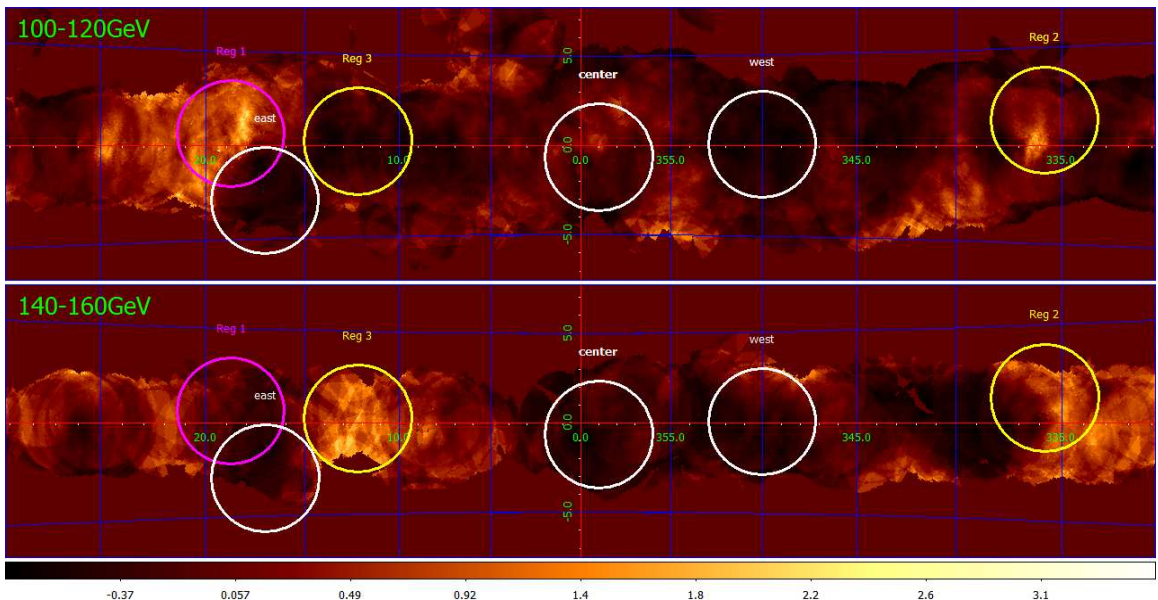}
\caption{Same as in Fig.~\protect\ref{fig:120_140GeV} but showing the
  difference of 100--120~GeV and 140--160~GeV energy bins with the half-sum of
  two corresponding adjacent bins.  Yellow and magenta circles mark bright excesses on
  these maps, see text for the details. The radii of all circles are $3^\circ$}
\label{fig:regions_description}
\end{figure*}

\bigskip

\section{Identifying regions with spectral features in the Galactic planes}
\label{sec:results}

For the analysis, presented below we used 209 weeks of \fermi\ data and
v9r23p1 Fermi Software. We filter the
photons with the expression {\small \tt (DATA\_QUAL==1) \&\& \\
  (LAT\_CONFIG==1) \&\& (ABS(ROCK\_ANGLE)<52) \&\& ((STOP<352773002.0) ||\\
  (START>352814402.0))} recommended by the \fermi-team in order to exclude a
bright solar flare that affected \fermi\ on March 8th, 2012.\footnote{ The
  large area telescope (LAT) on board the \fermi\ satellite is a
  pair-conversion gamma-ray detector operating between 20 MeV and 300 GeV. The
  LAT has a wide field of view of $\sim 2.4$~sr at 1 GeV, and observes the
  entire sky every two orbits ($\sim$3 hr for a \fermi\ orbit at an altitude
  of $\sim$565 km, full details of the instruments are given in
  \cite{Atwood09}).}

\subsection{Residual maps}
\label{sec:residual-maps}

We start by searching for regions around the Galactic plane ($|l| \le
30^\circ$, $|b| \le 5^\circ$) which have excesses at different energies. To
this end we split the energy range 80--200~GeV into 6 energy bins, 20~GeV each
(at energies of interest the spectral resolution of \fermi\ is about
$10\%$). In each energy bin we build a count maps using CLEAN class of photons
and smooth each map with a tophat filter with the radius of 3\deg, obtaining
images $\mathcal{I}_i, i=1\dots 6$. For each image $\mathcal{I}_i$ we then
define a ``background'' as
$\mathcal{B}_i=\frac{1}{2}\left(\mathcal{I}_{i-1}+\mathcal{I}_{i+1}\right)$. In
order to identify regions with spectral features we built then a set of images
$\mathcal{A}_i=\mathcal{I}_i/\mathcal{B}_i-1$.  Such a map for 120-140~GeV
energy band is shown in Fig.~\ref{fig:120_140GeV}. The positive (negative)
color coding at these images corresponds to the regions with positive
(negative) excess in the spectrum. Three regions with the most significant
excess from \cite{tempel12} are marked on the map as white circles.  The
positions of white circles obviously correspond to the positions of (some of)
the brightest excesses on this map. It is clearly seen in
Fig.~\ref{fig:120_140GeV} that emission around the GC region is shifted from
$(l,b) = (0,0)$.

Notice, however, that excesses of similar significance are visible not only on
120--140~GeV map (Fig.~\ref{fig:120_140GeV}), but also on the maps
100--120~GeV and 140--160~GeV (Fig.~\ref{fig:regions_description}). We
identified three additional regions that show high signal/background ratio:
\magenta\ and \second\ regions are bright on 100-120~GeV map, while \third\
region is bright on 140-160~GeV. All these regions are circles with 3\deg
radius, as are the regions of \cite{tempel12}.\footnote{ Any difference
  between the ``residual maps'' between CLEAN and ULTRACLEAN events would
  signify that these residuals are parts of some known systematics
  (e.g. contamination from cosmic rays)~\cite{Su-GC}.  We have verified that
  one obtains the same maps when using ULTRACLEAN photon class.}

The resulting spectra in the energy range 50--200~GeV extracted from the
regions \magenta, \second, and \third\ together with the \centerreg\ region
(coinciding with the region of the most significant excess of \cite{tempel12}
around the GC) are shown in Fig.~\ref{fig:spectra}.  The spectra from all the
regions \magenta, \second, and \third\ clearly show a number of features at
different energies. The thickness of the lines representing the spectra in
Fig.~\ref{fig:spectra} is defined by the formal $1\sigma$ statistical error.

\subsection{Analysis of the spectral features}
\label{sec:sign-feat}
\begin{small}
  \begin{table*}[t]
    \centering
    \begin{tabular}{lcccccc}
      \toprule
      Region  & $\chi^2_{50-200}$/28 & $\chi^2_{20-150}$/24 & $p_{50-200}$ &
      $p_{20-150}$ & $S_{50-200}$  [$\sigma$] & $S_{20-150}$ [$\sigma$] \\
      & (a) & (b) & (c) & (d) & (e) & (f) \\
      \midrule
      Reg 1 & 1.19 &1.64 &0.224 &2.5$\times 10^{-2}$ &1.22 & 2.24\\
      Reg 2 (shifted) & 1.21 &1.35 &0.205 &0.117 &1.27 & 1.57\\
      Reg 3 & 1.05&1.49  &0.392 &5.8$\times 10^{-2}$ &0.86 & 1.90\\
      GC    & 1.53 &1.78  &$3.6\times 10^{-2}$ & $1.1\times 10^{-2}$ &2.10 & 2.54\\
      \bottomrule
    \end{tabular}
    \caption{The values of $\chi^2/\mbox{[d.o.f.]}$ when fitting the power law
      model~(\protect\ref{eq:1}) to the data in  the energy range 50--200~GeV  (column
      (a)) and  20--150~GeV energy range  (column (b)). Columns (c) and (d)
      show the chance probability
      $p$ to get observed $\chi^2$ values (as defined by
      Eq.~(\protect\ref{eq:2})). The significances that correspond to these p-values
      are given in columns $S_{50-200}$ and $S_{20-150}$. The change of the
      fitting interval may affect the p-values by as much as an order of
      magnitude.
    }
    \label{tab:p_values}
  \end{table*}
\end{small}
\begin{small}
  \begin{table*}[!h]
    \centering
    \begin{tabular}{lllcl}
      \toprule
      & Prominent features,&
      Observed&Predicted&Significance \\
      Region (l$^\circ$, b$^\circ$) & energy range&  counts    & by model & of spectral feature \\
      \multicolumn{1}{c}{\textit{(a)}} &
      \multicolumn{1}{c}{\textit{(b)}} &
      \multicolumn{1}{c}{\textit{(c)}} &
      \multicolumn{1}{c}{\textit{(d)}} &
      \multicolumn{1}{c}{\textit{(e)}}
      \\
      \midrule
      \magenta\ (18.72$^\circ$, 0.57$^\circ$)    & 105--120~GeV  & 29 &10.46 & 4.63\\
      \second\ (335.76$^\circ$, 1.255$^\circ$)  & 70--90~GeV     & 75
      &40.95 & 4.72 \\
      \second\ (335.76, 1.255$^\circ$          & 95--110~GeV  & 32 &16.83  &
      3.23\\
      \third\ (12.22$^\circ$ ,0.20$^\circ$)    & 80--85~GeV  &         13 &5.25
      & 2.74  \\
      \centerreg\ (359$^\circ$,-0.7$^\circ$)         & 125--135~GeV   & 25 &3.53 & 7.34
      \\
      & 105-115~GeV  & 17 &5.99  & 3.58 \\
      \bottomrule
    \end{tabular}
    \caption{The most significant features in the regions \magenta, \second, \third,
      and \centerreg,  as well as the features in the region \second. We fit the
      background counts to the power law model in the whole 50--200~GeV energy
      range (unless otherwise  specified)  and determine the most
      prominent deviations from this simple background model and their
      significance  (column \textit{(e)}).
      This \emph{significance of the spectral feature} is
      defined as the Poisson
      probability to observe $N$ counts or more, provided
      that the spectral model predicts $\lambda$ counts (specified in the column
      \textit{(d)}).\protect\footnotemark}
    \label{tab:fit}
  \end{table*}
\footnotetext{This definition of significance is different from the one, used
  in~\cite{weniger12}. See the discussion in Sec.~\ref{sec:disc-concl}.}
\end{small}

\begin{table*}[t]
  \centering
  \begin{tabular}{lccc}
    \toprule
    Region &  $\chi^2_{20-150}$/23 & $\Delta \chi^2$ & Significance of a
    feature [$\sigma$]\\
    & {\small power law + Gaussian
      fit} & & \\
    \midrule
    \centerreg & 1.179 &15.6 &3.95\\
    \magenta & 1.339 & 8.6 &2.93\\
    \third & 1.375 & 4.1 &2.02\\
    \bottomrule
  \end{tabular}
  \caption{Change in the total $\chi^2$ when adding a Gaussian at 130~GeV (\centerreg),
    at 112.5~GeV (\magenta) and at 85~GeV (\third) as compared to the power law fits, shown in
    Table~\protect\ref{tab:p_values}. A formal significance of these features
    is defined as $\sqrt{\Delta \chi^2}$ (as we add only one degree of freedom
    when fitting a normalization of the Gaussian). The features in the region
    \second\ do not have a Gaussian shape and we do not show it in this table.}
  \label{tab:gauss}
\end{table*}

\begin{figure*}[!t]
  \subfloat[]{\label{fig:reg1}\includegraphics[width=0.45\textwidth]{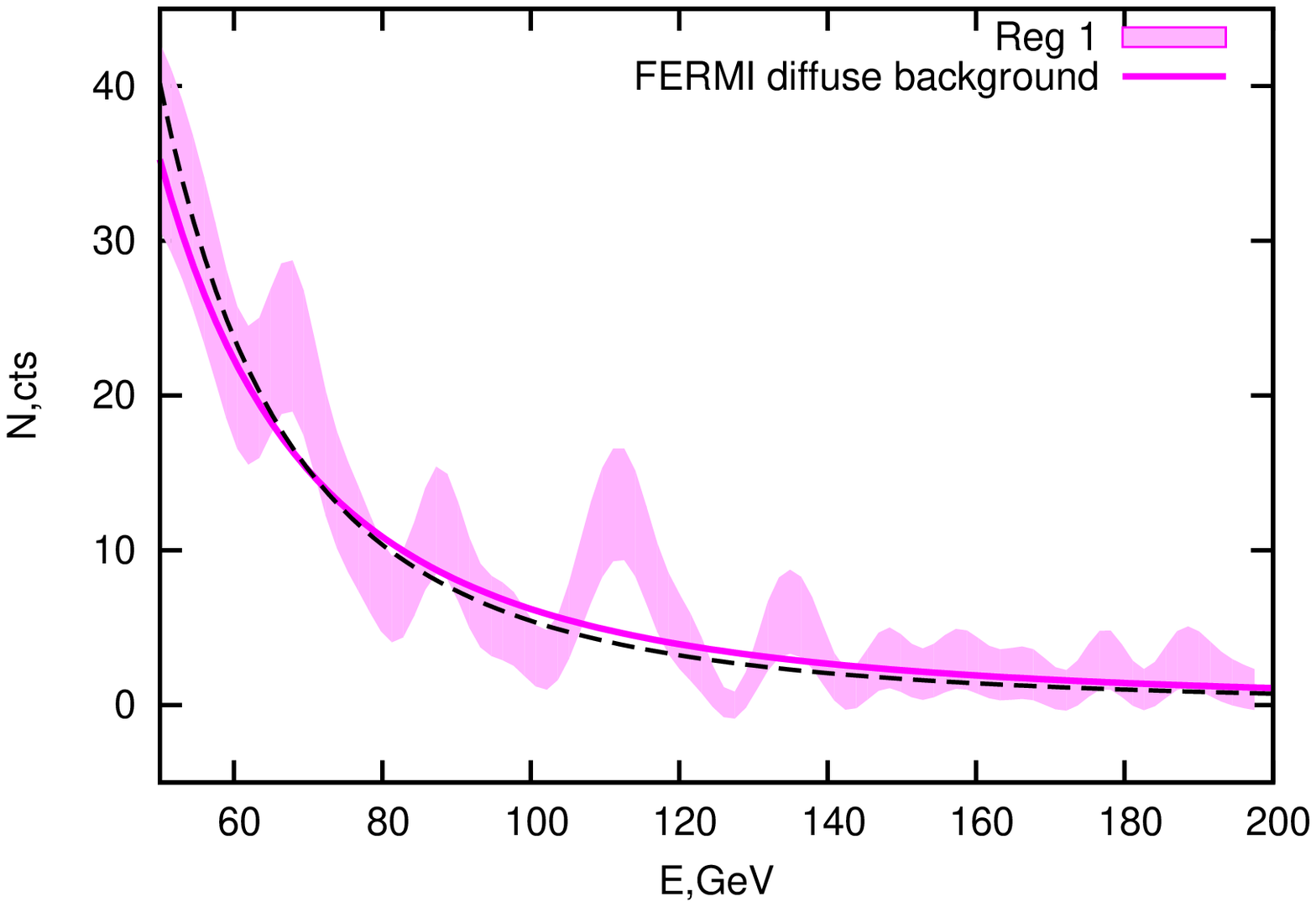}}
  \subfloat[]{\label{fig:reg2}\includegraphics[width=0.45\textwidth]{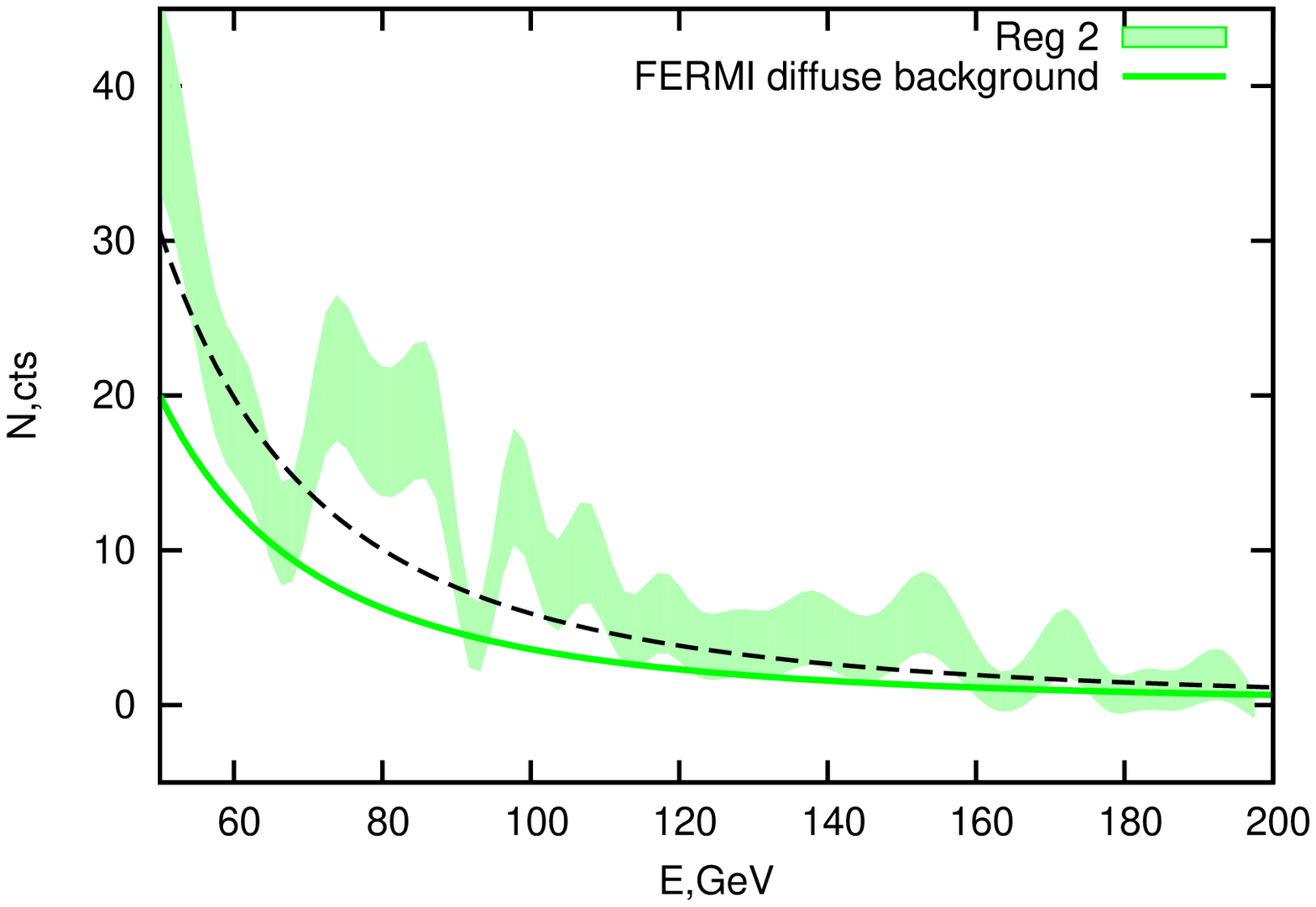}}\\
  \subfloat[]{\label{fig:reg3}\includegraphics[width=0.45\textwidth]{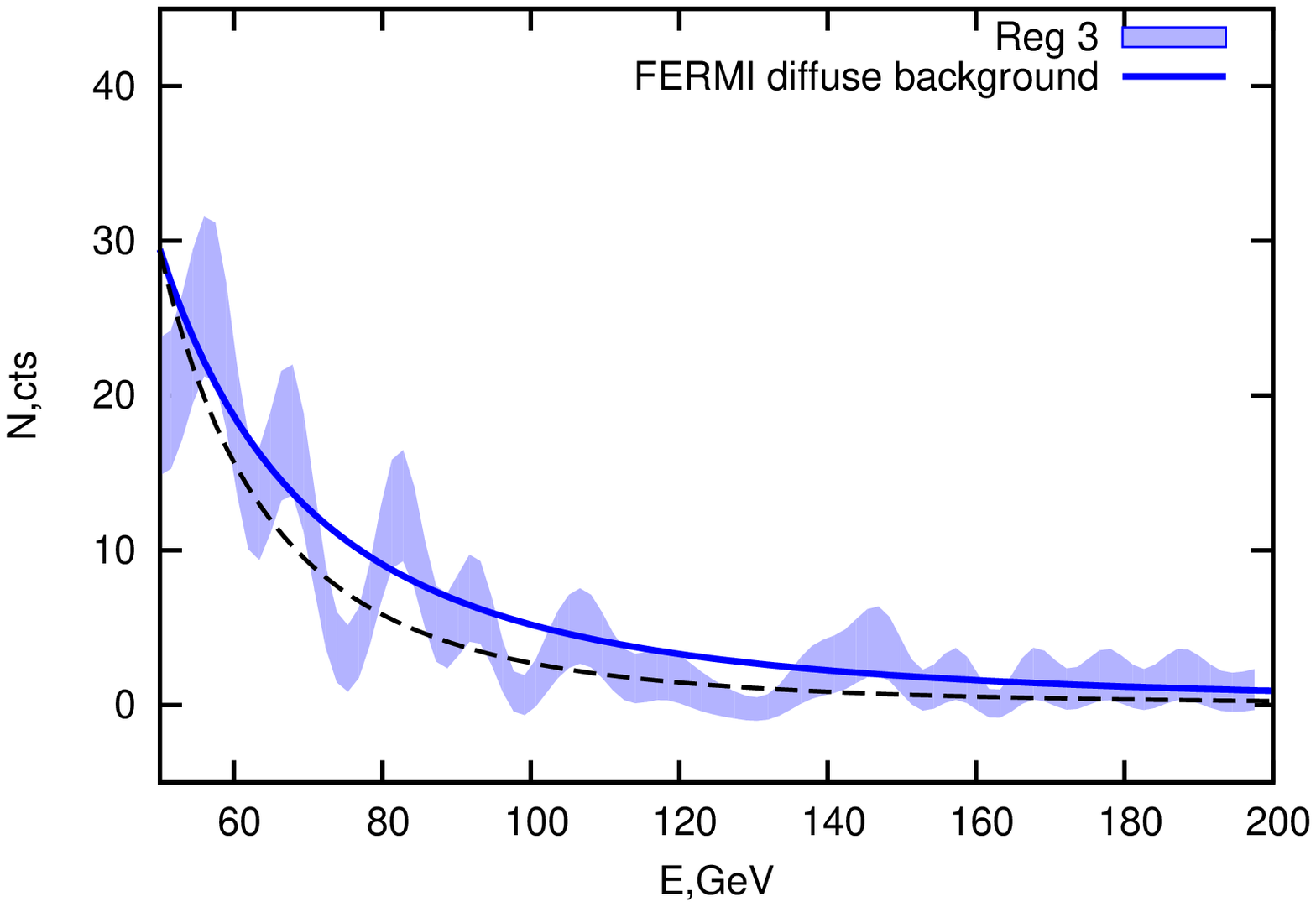}}
  \subfloat[]{\label{fig:center}\includegraphics[width=0.45\textwidth]{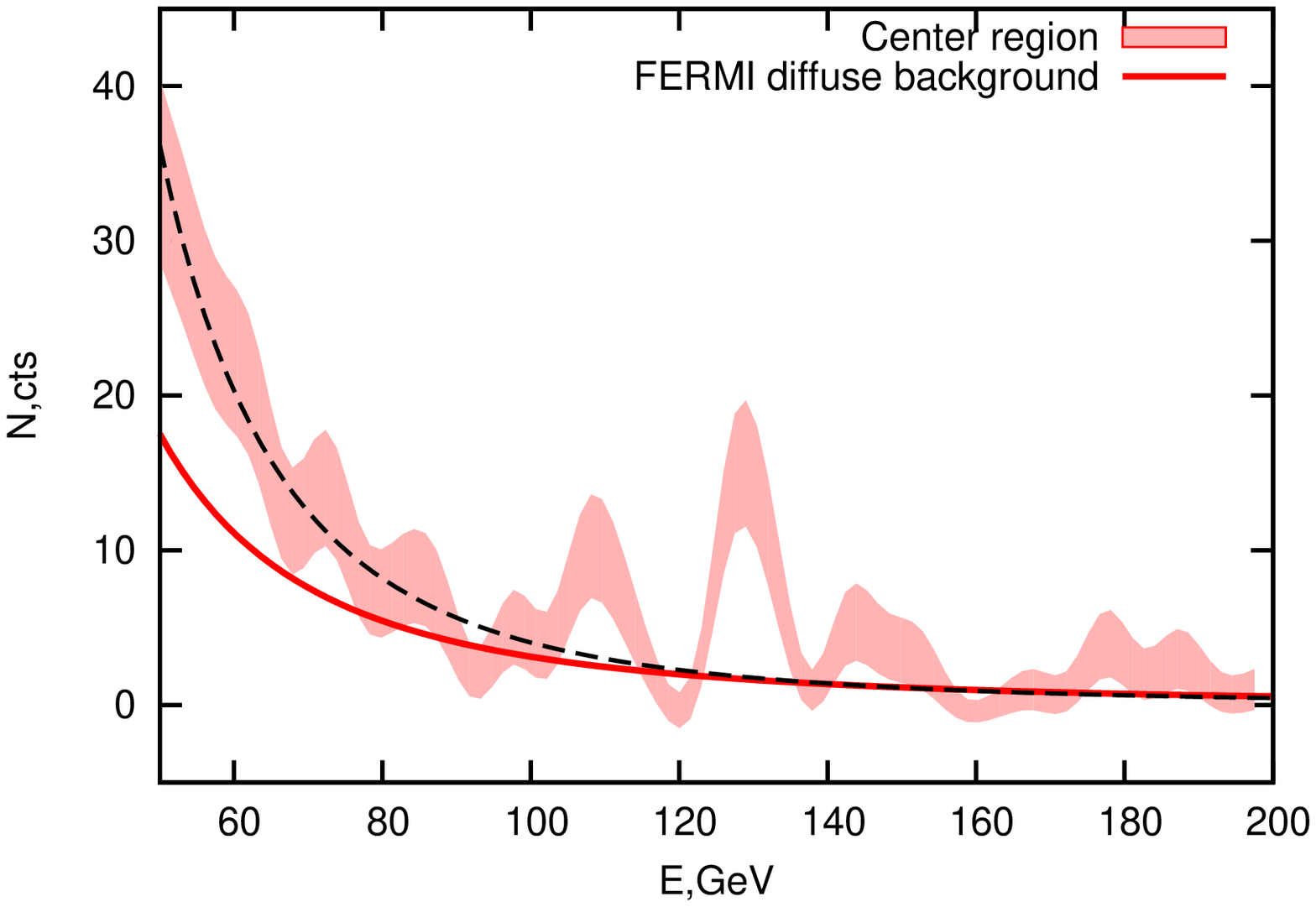}}\\
   \subfloat[]{\label{fig:allreg}\includegraphics[width=0.45\textwidth]{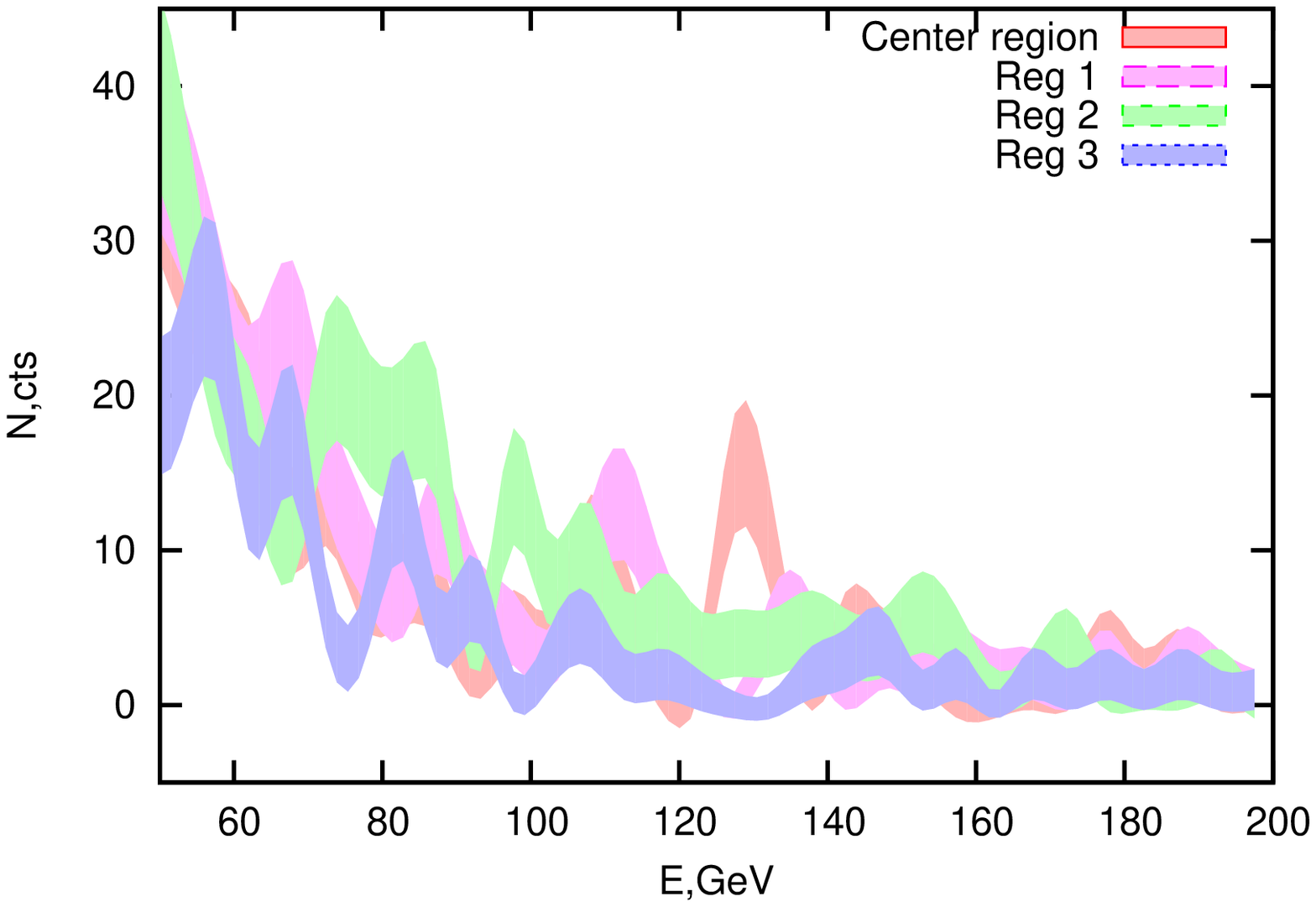}}
  \subfloat[]{\label{fig:reg2steps}\includegraphics[width=0.45\textwidth]{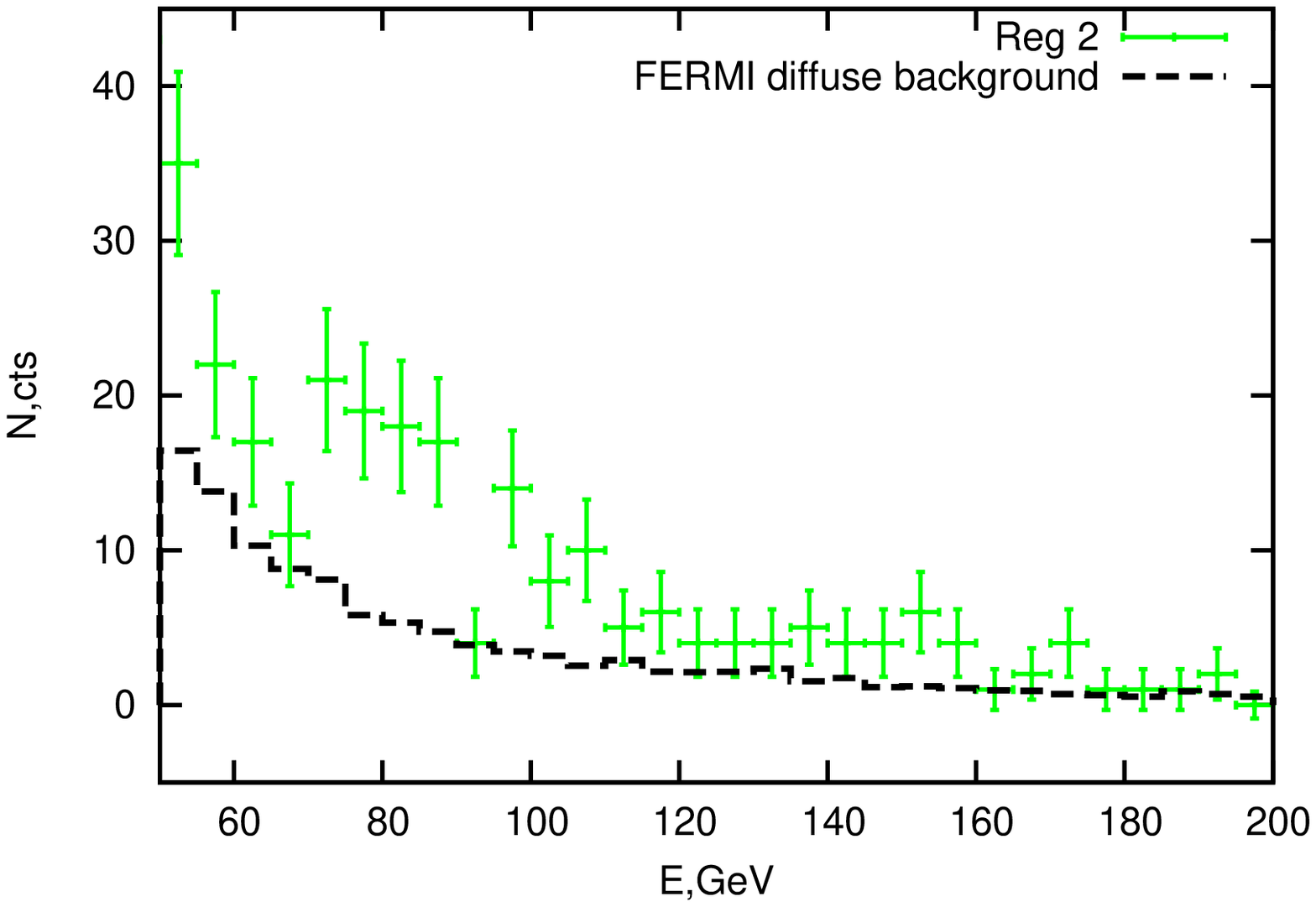}}
  \caption{Spectra from the Reg 1 region (magenta line),
    panel~\protect\subref{fig:reg1}), Reg 2 (panel~\protect\subref{fig:reg2}),
    \third\ (panel~\protect\subref{fig:reg3}) and central region (panel
    ~\protect\subref{fig:center}). The definitions of the regions are shown in
    Fig.~\protect\ref{fig:regions_description}.  The best fit power law for
    the central region is shown in black dashed line, the best fit diffuse
    FERMI background model is shown with solid color line.  The thickness of
    the spectra is determined as $\pm 1\sigma$ around the central value.  All
    spectra together are shown in panel~\protect\subref{fig:allreg}. The
    spectrum of Reg 2 with 10~GeV energy binning together with the
    corresponding prediction from the FERMI diffuse background model is given
    in panel~\protect\subref{fig:reg2steps} for comparison (see
    Section~\protect\ref{sec:tsmaps} for details).}
\label{fig:spectra}
\end{figure*}

In order to quantify the significance of the observed spectral features in the
regions \magenta, \second, \third\ we start with performing a power law fit
\begin{equation}
  N(E)=N_{100}(E/100~\mathrm{GeV})^{-\Gamma}, \label{eq:1}
\end{equation}
to the data. The reduced $\chi^2$ and the resulting p-values are presented in
the Table~\ref{tab:p_values}, columns (a--d). Here the p-values are defined as
\begin{equation}
  \label{eq:2}
  \mbox{p-value} = \int_{\chi_0^2}^{\infty}P_n(\chi^2) d\chi^2
\end{equation}
where $P_n(\chi^2)$ is the probability density function of the $\chi^2$
distribution with $n$ degrees of freedom.

Looking at the quality of fit to the power law model in $50$--$200$~GeV energy
range one could have concluded that with $20-40\%$ probability the data is
consistent with \emph{being described purely by the power law without any
  features} (even in the \centerreg\ region this probability is more than
3\%). Based on a similar analysis, it was argued (see e.g.~\cite{IDM,COSMO})
that the features in the regions \magenta, \second, \third are not
statistically significant.

However, as was discussed in the introduction, the relevant question is
whether one can demonstrate with confidence that the line is \emph{not}
present in the spectrum of the regions \magenta--\third and that the observed
photon counts are simply statistical fluctuations around a featureless
model. Below we argue that this is not the case.

We notice first of all, that the quality of fit (and therefore the conclusion
about ``chance probability'' of a given $\chi^2$) is sensitive to the
inclusion or omission of the bins with $E\ge 150$~GeV (as comparison of the
columns (a) and (b) in the Table~\ref{tab:p_values} demonstrates). These bins
have extremely low statistics (0 to 2 counts) and their large error bars
artificially improve the fit quality (as the comparison of the columns for
$\chi^2_{20-150}$ and $\chi^2_{50-200}$ demonstrates).  Inclusion of even
larger number of empty bins with $E>200$~GeV would further improve the quality
of fit.  At the same time, the change of the interval can change the resulting
p-values by as much as the order of magnitude (see the columns (c) and (d) in
the Table~\ref{tab:p_values}).

A way to determine whether a spectral feature is a fluctuation (that is at
least partially free from the above-described ambiguities) is to add an
additional component (Gaussian or other spectrally localized feature) and see
whether the quality of fit improves. In doing this, we see that for \magenta,
\third\ and \centerreg regions the quality of fit improves (by $2-3\sigma$).
The results are presented in Table~\ref{tab:gauss}.

\begin{figure*}[!t]
  \subfloat[\magenta. A feature at 110~GeV is clearly
  visible]{\label{fig:reg1-bins}\includegraphics[width=0.45\textwidth]{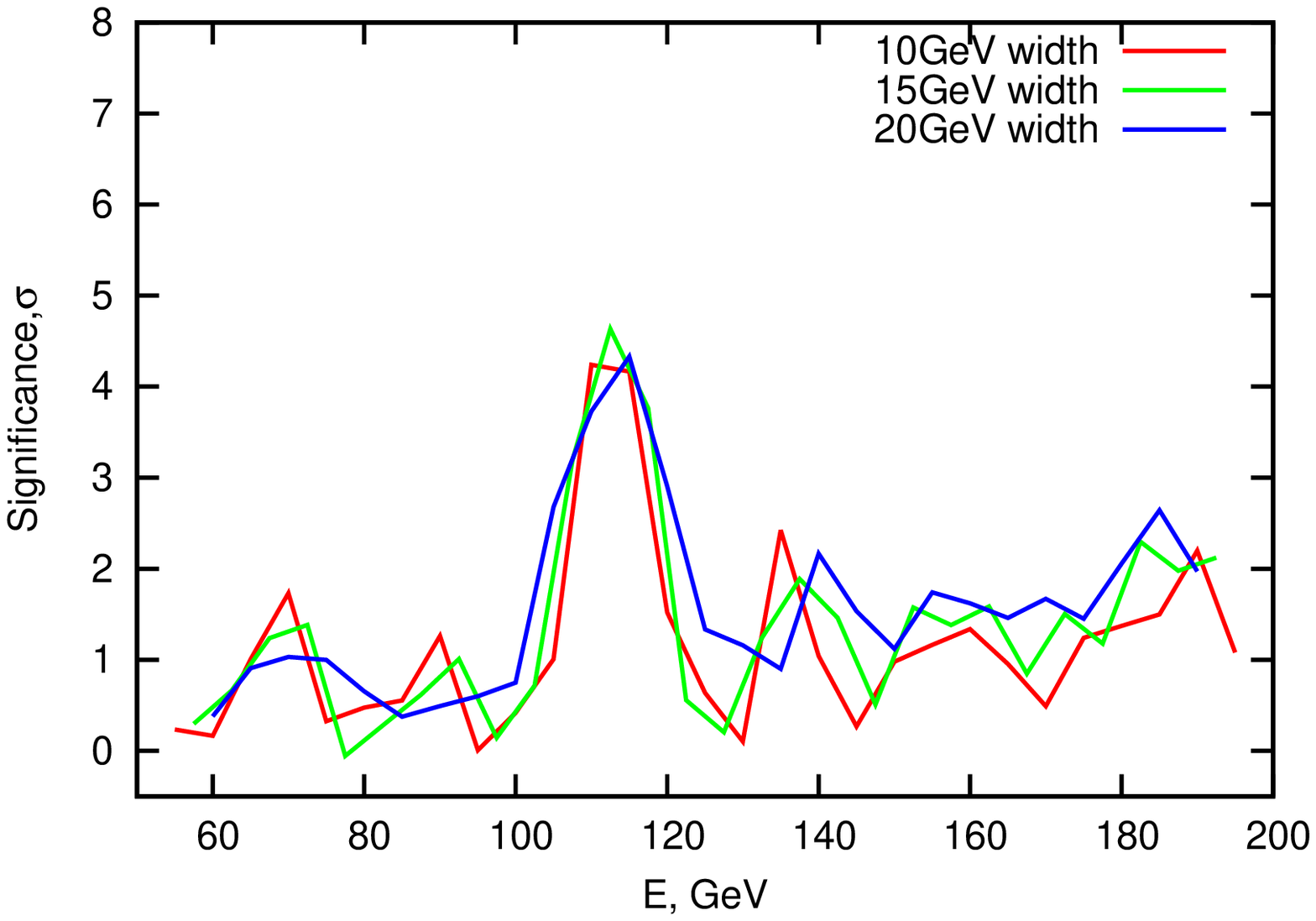}}~~%
  \subfloat[\second. A broad feature in the region $70-100$~GeV is
  clearly visible]{\label{fig:reg2-bins}\includegraphics[width=0.45\textwidth]{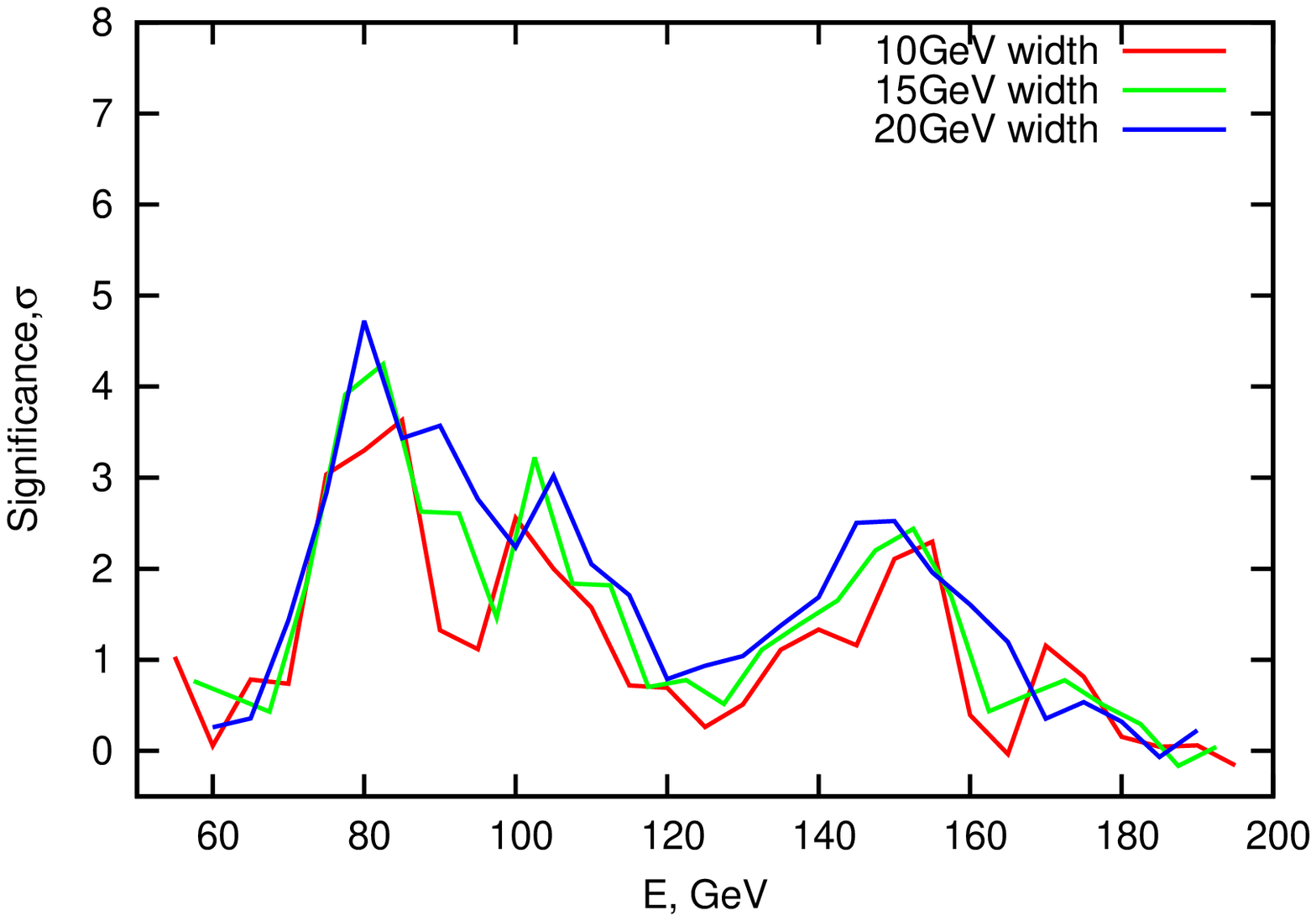}}\\
  \subfloat[\third. Possible features are around 80~GeV and around 150~GeV
  (compare excess in \second\ region at these energies). Notice that for this
  region we also added one bin local
  significance.]{\label{fig:reg3-bins}\includegraphics[width=0.45\textwidth]{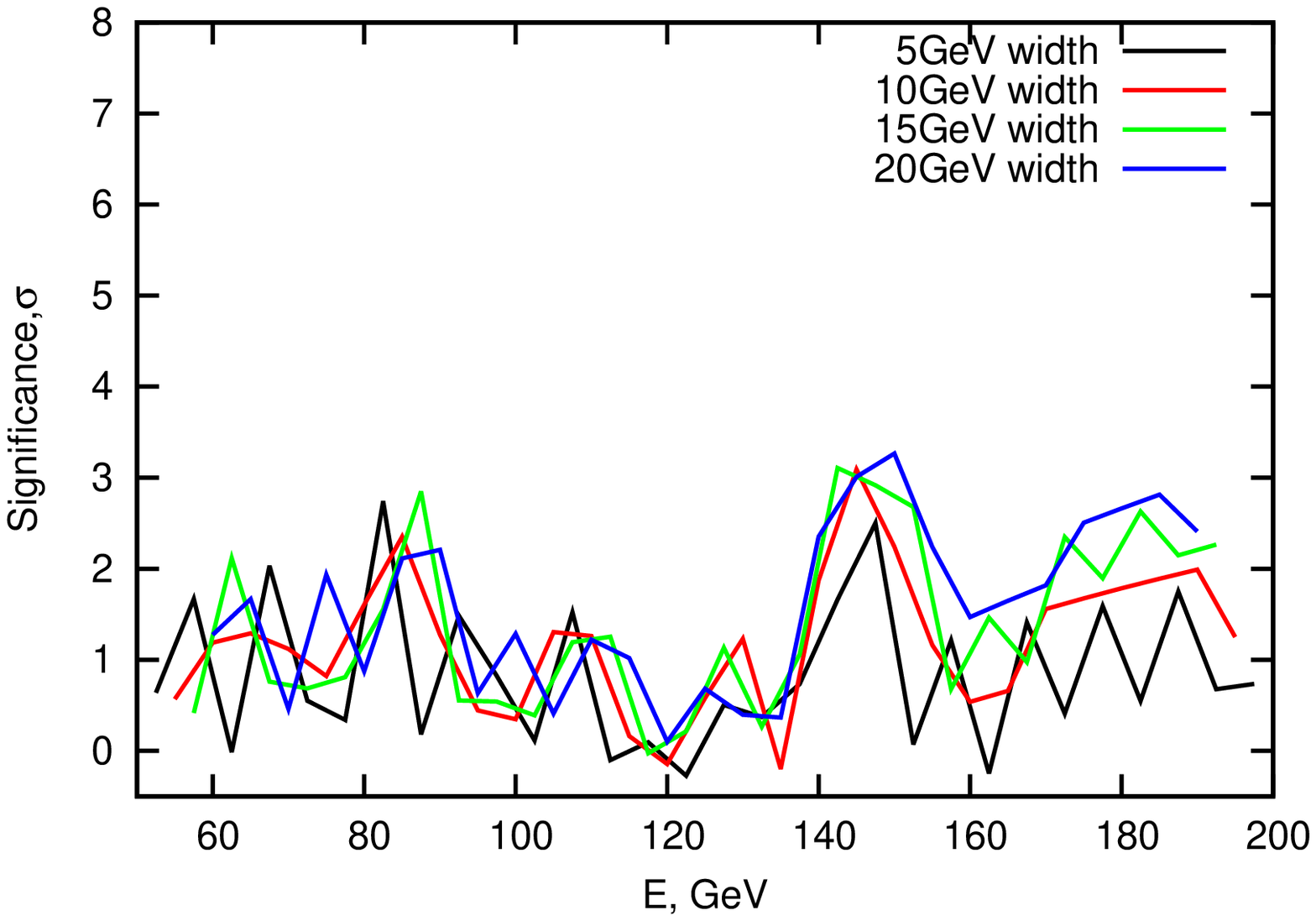}}~~
  \subfloat[\centerreg. A strong feature around 130~GeV and a possible second
  feature around
  110~GeV.]{\label{fig:center-bins}\includegraphics[width=0.45\textwidth]{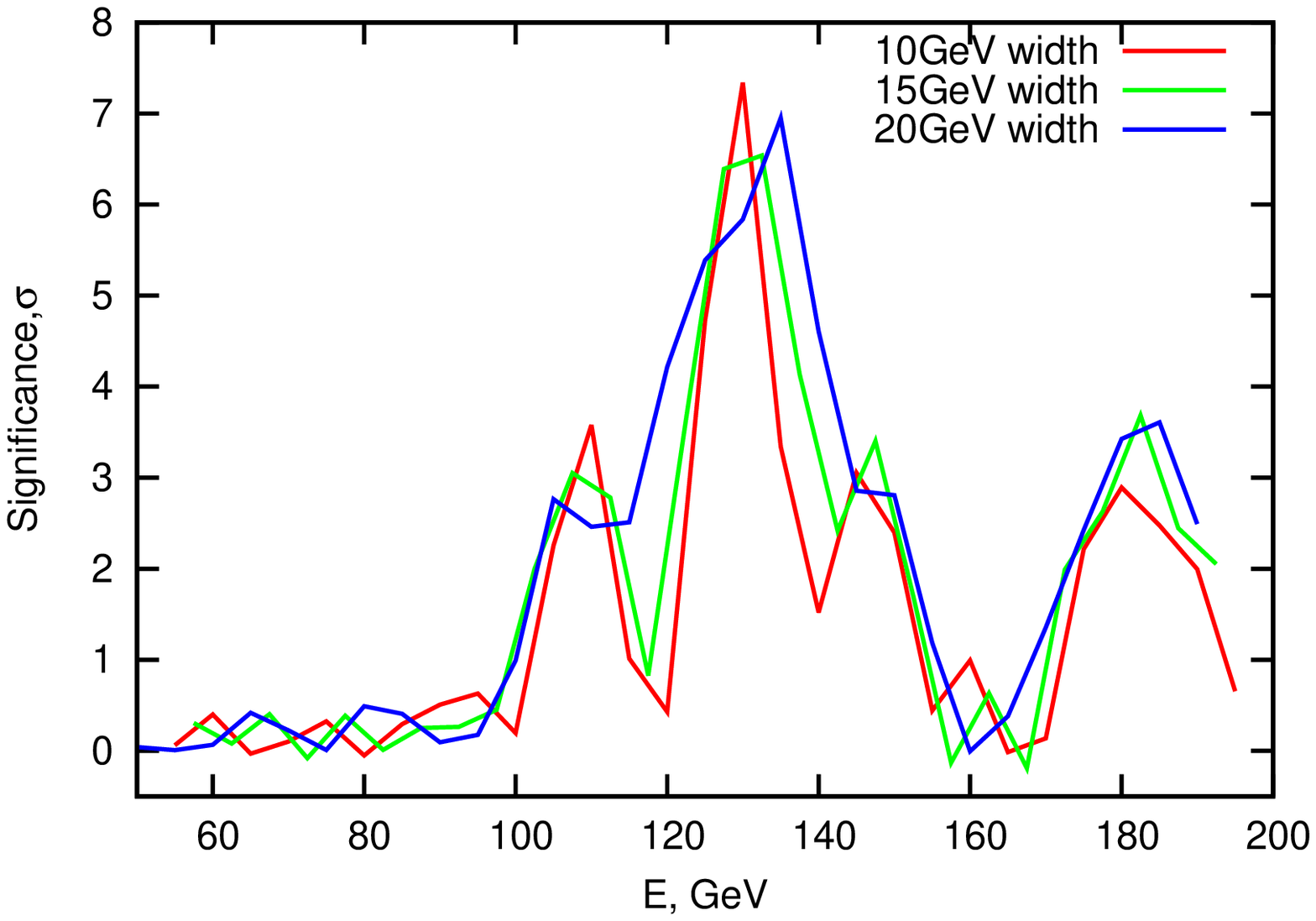}}
  \caption{Local significance of observing fluctuations in 2, 3 and 4
    consecutive bins (with 10, 15 and 20 GeV width correspondingly), as a
    function of energy. The significance is defined as the Poisson probability
    to observe $N$ or more counts, when the model predicts $\lambda$ (based on
    the power law background model). Notice that the significance does not
    depend on the number of consecutive bins that we choose!}
\label{fig:local-sign}
\end{figure*}

This method, however, crucially depends on our knowledge of a spectral shape
of a feature. As a result, we cannot characterize in this way, for example,
the region \second, where the feature at 70--100~GeV is clearly non-Gaussian.
Therefore we also compute the local significance of these features --- the
Poisson probability to observed $n$ counts or more in a given bin, provided
that the model predicts $\lambda$ counts. The results are presented in the
Table~\ref{tab:fit} and in Fig.~\ref{fig:local-sign}. Clearly, such a
\emph{local significance} should be higher than the p-value, presented in the
table~\ref{tab:p_values}. And indeed, in the spectrum of the \centerreg\
region we recover an excess at 130~GeV at $7.34\sigma$ and a $3.6\sigma$
excess in the energy bin around 110 GeV line.  The spectrum of the \magenta\
region demonstrates an excess above the power law fit at energy $\sim 115$~GeV
with the significance $4.63\sigma$ (Fig.~\ref{fig:reg1}). The spectrum of the
\second\ region (Fig.~\ref{fig:reg2}) formally has a reasonable power law fit
in the range 60--200~GeV (reduced $\chi^2=1.19$ for 28 degrees of freedom).
However, the distribution of residuals in this region shows a broad positive
fluctuations in the range 70--120~GeV.  Indeed, a power law fit to the subset
of the data (energies 60-110~GeV) is extremely poor (the reduced $\chi^2 >
2$).  However, in the region 110--200~GeV the data points have just few counts
and therefore large statistical errors reduce the overall $\chi^2$ value to
acceptable values close to 1.  Finally, in the region \third\ we observed a
line-like feature at 80~GeV (and a possible additional feature around
140--150~GeV) with significance around $3\sigma$. Notice that positions of
both features in \third\ coincide with the peaks in the local significance of
the \second. We also checked that thus determined significance does not depend
on the number of bins (the maxima of the lines of different color in the
panels of Fig.~\ref{fig:local-sign} coincide). Notice, that the bin of 20~GeV
is much wider than the spectral resolution of the \fermi satellite at these
energies (about 10~GeV), suggesting that this is \emph{not} a line-like
feature (whose width should be 10\% of the energy) but rather a power law with
a sharp cut-off (as e.g. in~\cite{Aharonian:2012cs}).

\begin{figure*}[!t]
  \centering %
  \subfloat[Region \centerreg. Result of $10^4$ Monte Carlo realizations, of
  which 99.9\% lie between dashed lines. Significance of the feature at
  130~GeV is \emph{more than}
  $3\sigma$.\label{fig:mc-gc}]{\includegraphics[width=.32\textwidth]{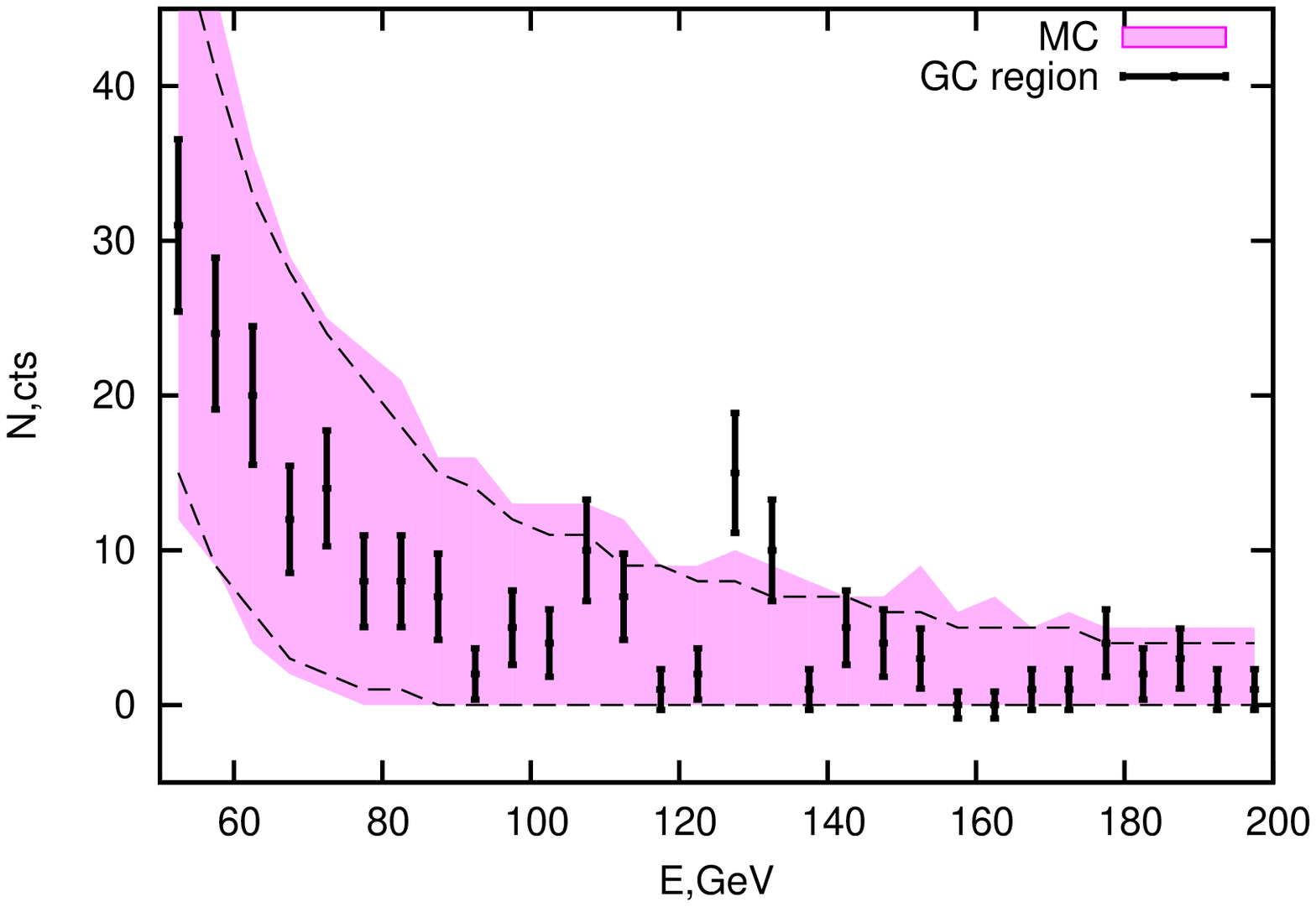}}~~
  \subfloat[Region \magenta . Result of $10^4$ Monte Carlo realizations, of
  which 99.9\% lie between dashed lines. Significance of the feature at
  110~GeV is \emph{more than}
  $3\sigma$.\label{fig:mc-reg1}]{\includegraphics[width=.32\textwidth]{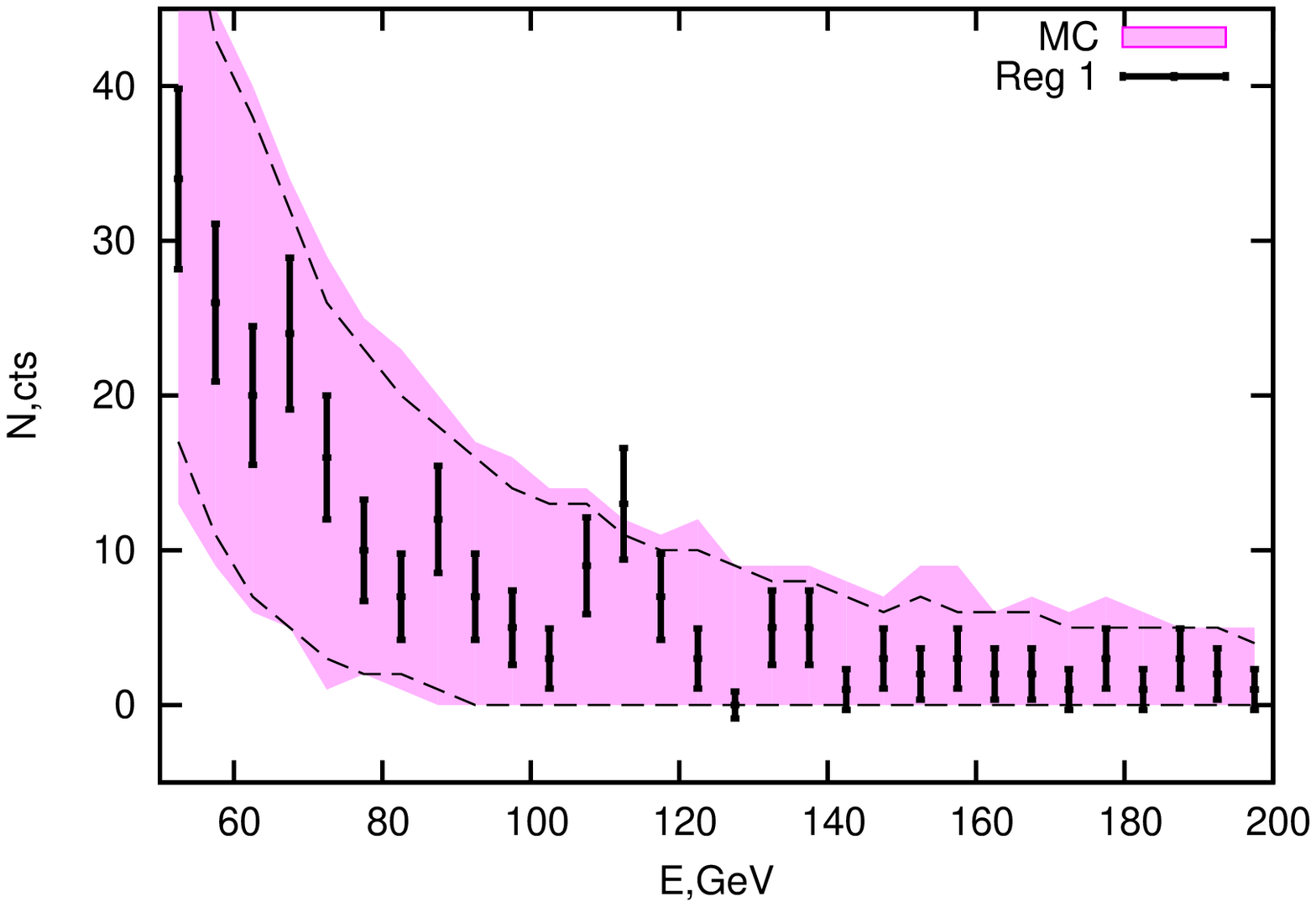}}~~
   \subfloat[Region \second. Result of $10^4$ Monte Carlo realizations, of
  which \textbf{99\%} lie between \emph{dashed-dotted lines}. Significance of
  each deviation is \emph{above}
  $2.3\sigma$.\label{fig:mc-reg2-99}]{\includegraphics[width=.32\textwidth]{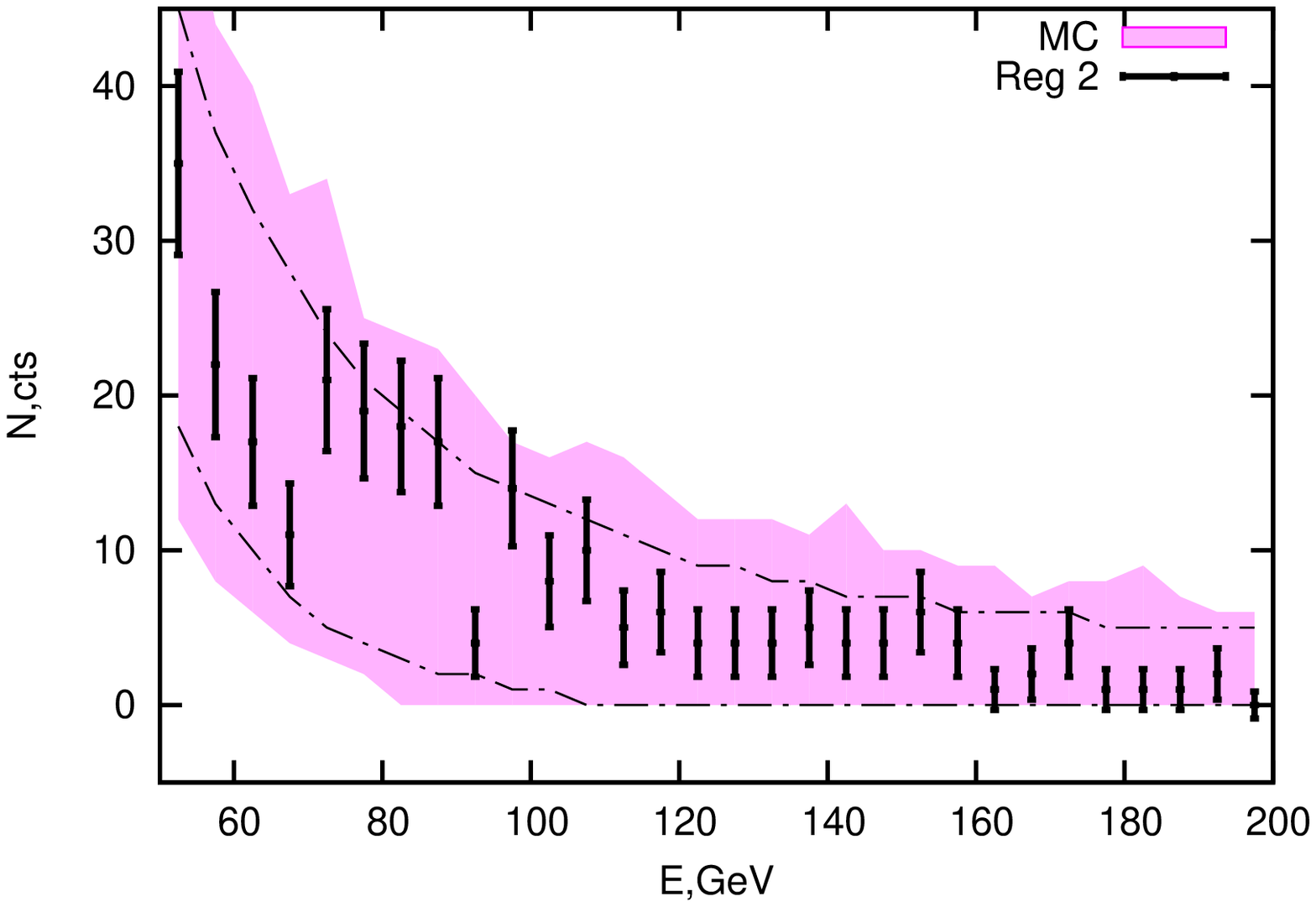}}
  \caption{Monte Carlo realizations of the power law models in the region
    50--200~GeV for \centerreg\ (left), \magenta\ (center) and \second\
    (right) regions. Number of realization is $10^4$. Shaded regions show
    minimum and maximum number of counts in all the Monte Carlo
    realizations. $99.9\%$ of all realizations lie \emph{between} dashed lines
    (in panel~\protect\subref{fig:mc-reg2-99} dashed-dotted line show 99\% of
    all models). Data points outside dashed lines are \emph{at least}
    $3\sigma$ deviations from the power law model. Two data points in the
    panel~\protect\subref{fig:mc-reg2-99} deviate from the power law by
    $2.3\sigma$.}
   \label{fig:mc}
\end{figure*}

\begin{figure*}[t]
  \centering
  \includegraphics[width=.45\textwidth]{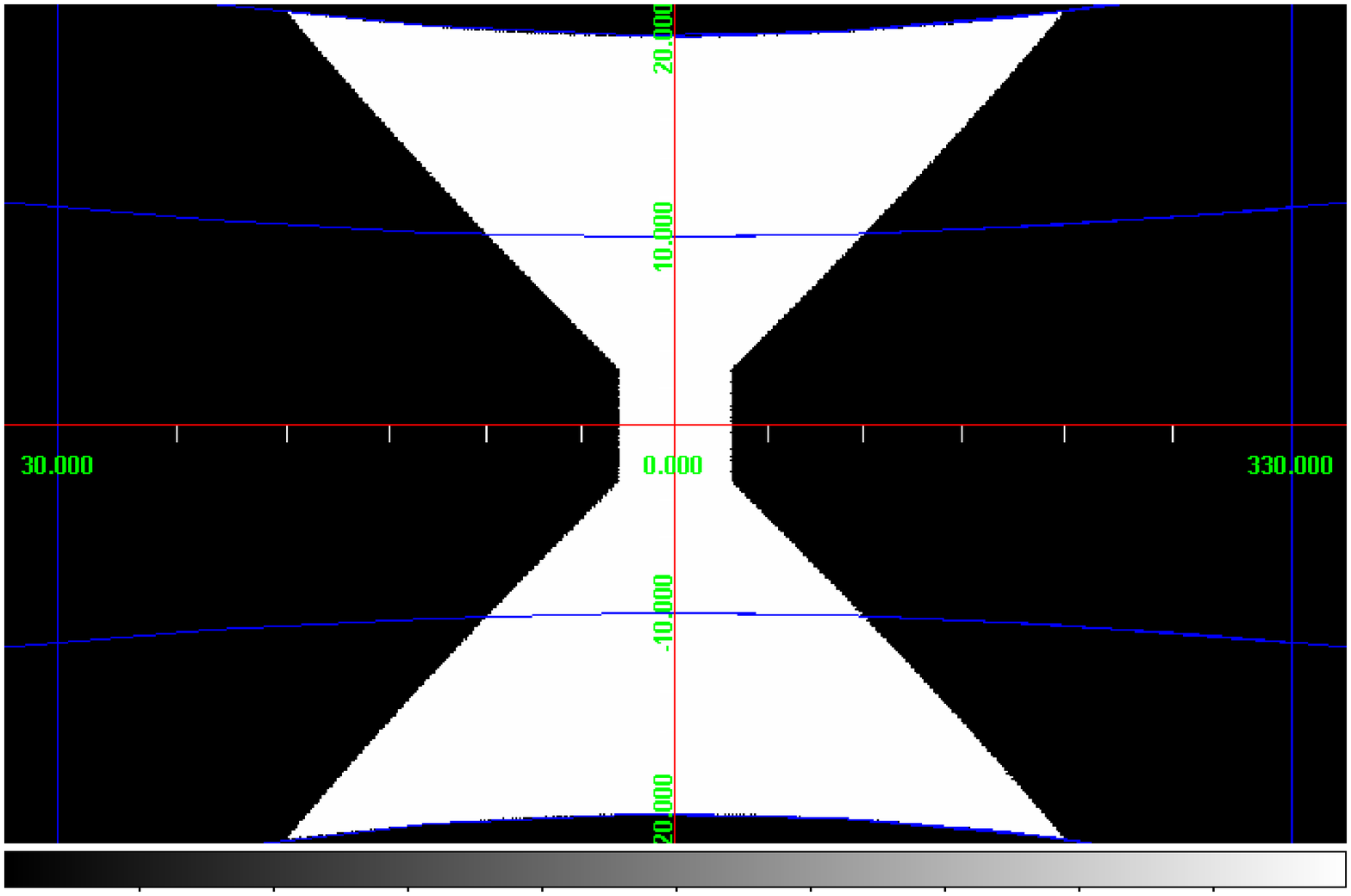}
  \includegraphics[width=.45\textwidth]{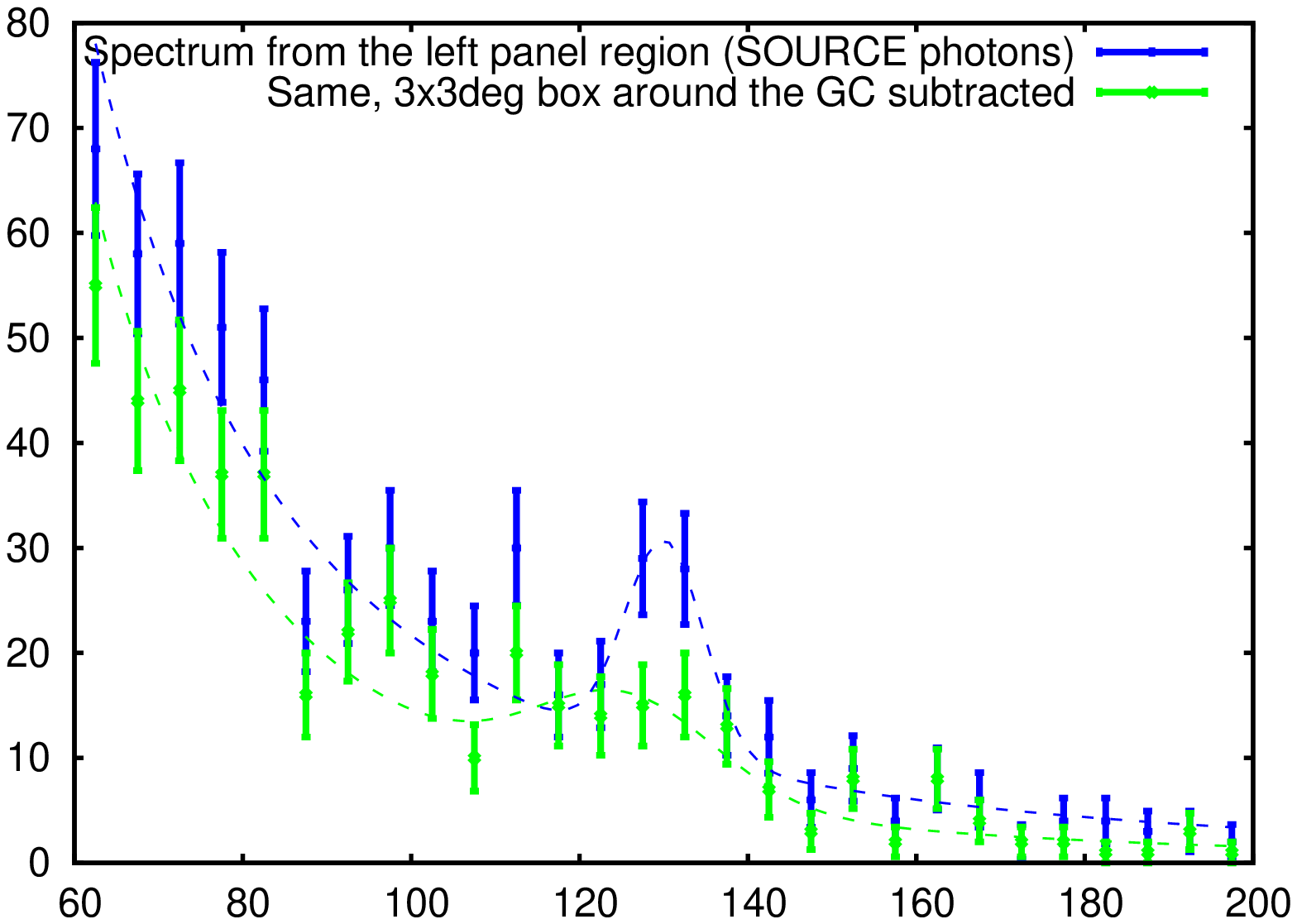}
  \caption{\textbf{Left panel:} region that approximately coincides with the
    ``Region 3'' of \protect\cite{weniger12} and reproduces its
    spectrum. \textbf{Right panel:} Spectrum of this region (blue) with the
    feature at 130~GeV clearly visible. The same spectrum extracted from the
    region with removed central $3\deg\times 3\deg$ box (green) and decreased
    significance of 130~GeV feature.}
  \label{fig:weniger}
\end{figure*}

In order to further clarify the significance of the observed features (and
correct for possible ``trial factors''), we perform $10^4$ Monte Carlo
realization of the background (power law) model for regions \magenta, \second\
and \centerreg (see Fig.~\ref{fig:mc}). In all panels 99.9\% of all
realizations are inside the dashed lines.  Feature at $130$~GeV in the
\centerreg\ region (Fig.~\ref{fig:mc-gc}) lies clearly outside these lines,
which means that the significance of the feature is \emph{above}
$3\sigma$. Feature at 110~GeV in \magenta (Fig.~\ref{fig:mc-reg1}) is also a
$3\sigma$ deviation.\footnote{The discrepancy between the p-value, deduced
  from Monte Carlo simulations, and p-values, listed in the
  Table~\ref{tab:p_values}, is of course not surprising --- $\chi^2$ is a
  global measure of the quality of fit, while the features that we are
  discussing are localized to several consecutive bins.}  Notice, that in case
of the \second\ our simulations provide a conservative estimate of
significance. The described procedure does not take into account that in the
spectrum of this region \emph{several consecutive bins} deviate from the power
law model.
This analysis allows us to conclude that, although the spectral feature at the
GC is more significant (possibly due to the lower \emph{predicted} background
at 130~GeV as compared to lower energies), \emph{the presence of the spectral
  features in other regions cannot be ruled out with confidence}.\footnote{The
  regions \magenta, \second, \third, discussed so far, as well as the regions
  where the most significant excess is observed in~\cite{tempel12}, are
  located in the different regions along the Galactic plane. In
  \cite{weniger12}, however, the excess at 130 GeV was claimed to originate
  from a large region, mostly located outside the Galactic plane (its
  approximate shape is shown in Fig.~\ref{fig:weniger}, left panel).  Removing
  the central box of $3\deg\times 3\deg$ (about $10\%$ of the total field of
  view) we see that the significance of the feature at 130~GeV drops (green
  vs.\ blue data points in the right panel in
  Fig.~\ref{fig:weniger}). Therefore, we conclude that the feature at 130~GeV
  is related to the region close to the Galactic center, without inclusion of
  this region the feature becomes insignificant (a similar conclusion was
  reached in \cite{tempel12,Su-GC}).}

\subsection{Test statistics maps based on \fermi diffuse background model}
\label{sec:tsmaps}

One possible interpretation of the above results is that by exploring regions
outside the GC one increases the ``trial factors'' by $10^\circ \times
60^\circ/(\pi \times 3^\circ \times 3^\circ) \approx 21$ (and therefore
p-vales of the features, obtained above, should be further penalized by this
factor). This would be the case if spectra in all $3^\circ$ circles around the
Galactic plane were completely arbitrary, not correlated with each other.
This is however, not true in our case, as the diffuse emission in the Galactic
Plane is described by the smooth model.

\begin{figure*}[!t]
  \centering \subfloat[Energy range 120--140 GeV. The maximal TS $\approx$21.2
  is located at the pixel with $(l,b)\approx
  (-1.5^\circ,-0.75^\circ)$.]{\label{fig:ts130}\includegraphics[width=.8\textwidth]{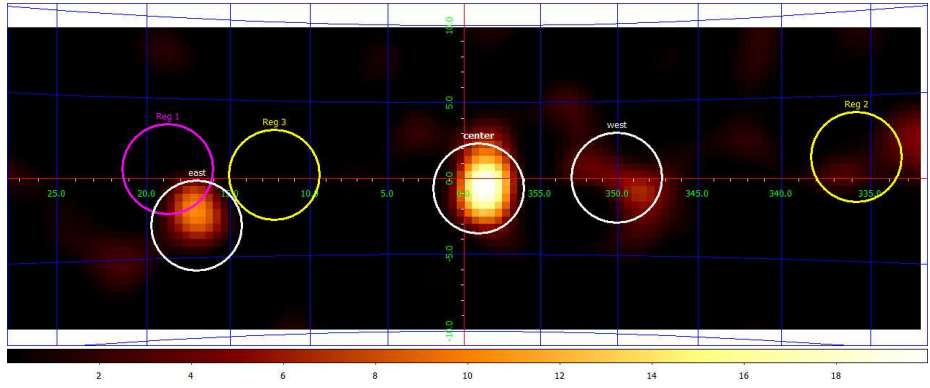}}\\
  \subfloat[Energy range 100--120 GeV. The maximal TS in \magenta\ is
  TS$\approx 15.0$ is located at the pixel with $(l,b)\approx (18.1, 0.2)$ and
  in the region \second\ the maximal TS is $\approx 17.3$, located at the
  position $(l,b) = (337.1, -0.2)$]{\label{fig:ts110}\includegraphics[width=.8\textwidth]{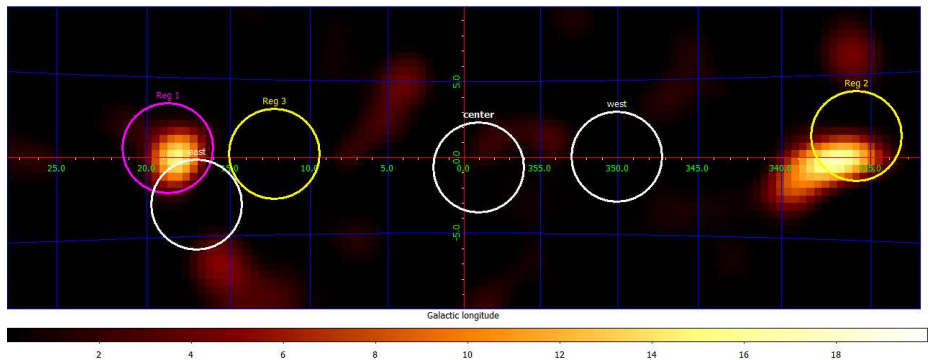}}\\
  \subfloat[Energy range 70--90 GeV. The maximal TS $\approx$32.3 is located
  at the pixel with $(l,b)\approx (336.0, -0.2)$.]{\label{fig:ts80}
    \includegraphics[width=.8\textwidth]{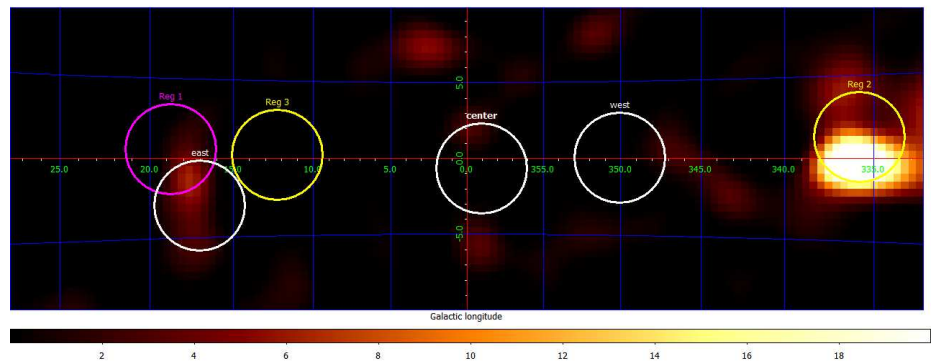}}
  \caption{Test-statistics (TS) maps when in addition to Galactic and
    extragalactic \fermi\ backgrounds we add a model of the $3^\circ$-region
    centered at this point. The latter was taken to has Gaussian spatial
    profile with $3^\circ$ FWHM and Gaussian spectral profile with $5$~GeV
    dispersion, centered correspondingly at 80~GeV, 110~GeV, 130~GeV.}
  \label{fig:tsmaps}
\end{figure*}

To study both spectral and spatial variations of emission around the Galactic
plane, we model $50^\circ\times 50^\circ$ region (in J2000 coordinates) around
the GC by three components --- standard galactic(centered at GC) and
extragalactic \fermi\ backgrounds\footnote{Given by
  \texttt{gal\_2yearp7v6\_v0.fits} and \texttt{iso\_p7v6clean.txt} templates,
  see e.g.\
  \url{http://fermi.gsfc.nasa.gov/ssc/data/access/lat/BackgroundModels.html}.}
and a model of a ``line'', that we try to add in each pixel in order to
improve the fit (pixel size is $0.5^\circ \times 0.5^\circ$). The ``line''
model has Gaussian spatial profile with $3^\circ$ FWHM and Gaussian spectral
profile with $5$~GeV dispersion, centered correspondingly at 80~GeV, 110~GeV,
130~GeV.\footnote{We have repeated the same exercise with the Gaussian line,
  having $2$~GeV FWHM and obtained similar results.} We fit the data with the
diffuse models and then consider the change of the log-likelihoods of the fit
when adding the Gaussian component at these three energies.  We build the map
of TS values (the difference of log-likelihoods of two models) using
\texttt{gttsmap} tool.  The normalizations of the \fermi\ diffuse backgrounds,
as well the normalization of the region model were allowed to vary during the
fit procedure. The test-statistics values of the $3^\circ$-region model fits
are shown in Figs.~\ref{fig:tsmaps}. The significance of the model can be
estimated as $\sqrt{TS}\sigma$.

The test-statistical value of the line centered at \centerreg\ region is about
21, that corresponds to significance $\sim 4.5\sigma$. Beside the \centerreg\
region Figs.~\ref{fig:tsmaps} exhibit additional regions with the excesses at
70--90~GeV (Fig.~\ref{fig:ts80}) and 100-120~GeV
(Fig.~\ref{fig:ts110}). Notice that, contrary to the results of~\cite{Su-GC},
the test statistics of the \centerreg\ region does not improve significantly
when adding a Gaussian at 110~GeV, i.e. photon distribution in this energy bin
strongly deviates from $3^\circ$ FWHM spatial Gaussian profile.  The highest
test statistics $\mbox{TS} = 32.3$ is found in the region \second\ at $80$~GeV
(see Table~\ref{tab:gauss} for details).

  \begin{table*}[t]
    \centering
    \begin{tabular}{llr}
      \toprule
      Region ($l^\circ$, $b^\circ$) &  Energy of  the line & $\mbox{TS}_{max}$  \\
      \otoprule
      \magenta\ (18.72$^\circ$, 0.57$^\circ$)    & 110~GeV & 14.99\\
      \lightmidrule
      \second\ (335.76$^\circ$, 1.255$^\circ$)  & 80~GeV  & 32.3 \\
      & 110~GeV & 17.26\\
      \lightmidrule
      \third\ (12.22$^\circ$ ,0.20$^\circ$)    & 80~GeV  & 2.74 \\
      \lightmidrule
      \centerreg\ (359$^\circ$,-0.7$^\circ$)       & 130~GeV & 21.22\\
      &    110~GeV & 4.5\\
      \bottomrule
    \end{tabular}
    \caption{Change of the \emph{test statistics} (TS)
      when fitting spectral and spatial distribution of photons to a smooth
      background model and to the same model with an additional line (see Section~\ref{sec:results})
      The line  has Gaussian spatial profile with $3^\circ$ FWHM and Gaussian spectral
      profile with $5$~GeV dispersion on top of \fermi diffuse background
      models. The TS maps are shown in Fig.~\ref{fig:tsmaps}.
      The \emph{$TS_{max}$ column} shows the maximum difference of likelihoods
      over all pixels within a given region.}
    \label{tab:TS}
  \end{table*}

\section{Discussion and conclusions}
\label{sec:disc-concl}

The first version of this paper
(\href{http://arxiv.org/abs/arXiv:1205.4700v1}{1205.4700v1}) demonstrated that
spectra of several regions (e.g.\ \magenta, \second, \third) are not
featureless at energies $E > 50$~GeV, but contain excesses similar in
significance to the spectral feature in the GC region.  In particular,
version~1 found excess in $110$~GeV bin from the \magenta (a pair of lines at
$\sim 110$ and $\sim 130$~GeV from the GC region was later discussed
in~\cite{Su-GC}).  Our present analysis recovers the previous results with
high significance and our main conclusion remains intact -- with the current
data it is impossible to \emph{rule out the possibility} that the observed
spectral features are actual signals (of astrophysical or instrumental
origin), rather than statistical fluctuations of the smooth
backgrounds. Clearly, until such a possibility is ruled out, the DM
interpretation of the emission from the GC remains dubious.

Refs.~\cite{Su-GC,Finkbeiner:12a} also analyzed \emph{spatial} distribution of
residuals and significance of different regions around the Galactic plane.
The analysis, presented here, and the interpretation of the significance of
the spectral features is different due to the different statistical approaches
adopted.

One way to detect spacial ``hot spots'' with significant spectral features is
just to look for the regions where the best fit to a smooth (power law) model
is unacceptable (i.e.\ the reduced $\chi^2\gg 1$). This method was discussed
e.g.\ in~\cite{COSMO,IDM}. The results of such fits for our regions were
presented in Table~\ref{tab:p_values}. This method however, is prone to a
number of uncertainties, related to the choice of fitting interval and to the
fact that $\chi^2$ is a global measure of fit, where the significance of local
spectral features can be ``blurred'' by many bins with large errors
bars. Indeed, as demonstrated in Section~\ref{sec:sign-feat} the \emph{local
  significance} of the features, found in regions \magenta, \second, and
\third\ is higher than predicted by this method ($3\sigma$ and above).

\begin{figure*}[t]
  \centering
  \includegraphics[width=.5\textwidth]{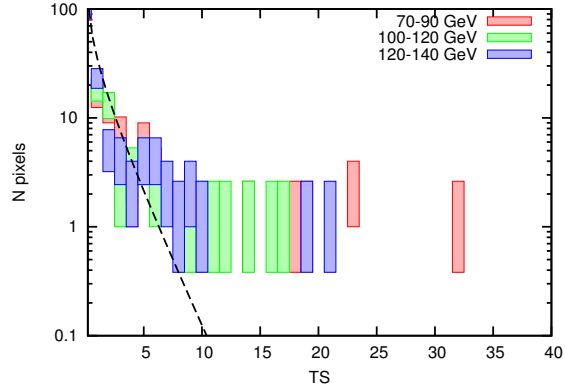}
  \caption{Distribution of \emph{test statistics} (TS) when adding a fixed
    width (in space and in energy) Gaussian in different positions in the
    Galactic plane (as described in Section~\ref{sec:results}, see also
    Figs.~\ref{fig:ts130}--~\ref{fig:ts80}). Dashed line is the $\chi^2$
    distribution with 1 degree of freedom.  If all pixels with TS${}> 0$ in
    Figs.~\ref{fig:tsmaps} are merely statistical fluctuations, one should see
    that the number of pixels for each TS is consistent with the dashed line
    and as a result there are not pixels with TS${}\gtrsim 8$ (for the number
    of pixels that we have in our regions).  However, if high-TS outliers for
    \centerreg, \second, \magenta, etc. are not caused by statistical
    fluctuations only, one should see a tail of large TS (as indeed seen in
    this Figure).}
  \label{fig:deltaTS}
\end{figure*}

A common way to determine the significance of these hot spots is to multiply
the resulting p-values of the spectral features by the number of regions that
were used in the search for the ``hot spots'' (so called ``trial factors'',
see e.g.~\cite{Su-GC}). The number of regions is, however, totally subjective
and the resulting significance can be artificially lowered (or increased). We
believe that a better way to account for spatial distribution of residuals is
to find a smooth background model that fits the data in (most of the) spatial
regions and deviation from which are normally distributed (modulo some
localized spots) (a similar approach was used in~\cite{Su-GC}).  We used the
galactic diffuse background by the \fermi\ collaboration as such a spatial
model. We see (Figs.~\ref{fig:tsmaps}) that this model correctly fits the data
``almost everywhere'' apart from several localized spots where it
underpredicts the number of photons and where additional Gaussian strongly
improves the quality of fit.  The histogram (Fig.~\ref{fig:deltaTS}) clearly
shows that the distribution of pixels with TS${}\lesssim 8$ is consistent with
the $\chi^2$ distribution with 1 degree of freedom (as it should be for
statistical fluctuations). However, there are also few \emph{high TS}
outliers, corresponding to the hot spots in
Figs.~\ref{fig:ts130}--\ref{fig:ts80}).\footnote{There are about 133
  independent $3\deg \times 3\deg$ pixels in the region ($|l|\le 30$\deg,
  $|b|\le 10$\deg) that we used for this analysis. If all positive TS are
  caused just by statistical fluctuations, one expects (roughly) one outlier
  with $\mbox{TS}\sim 9$ for $\sim 300$ pixels.} These regions are the GC and
several other regions (\magenta\ and \second) that we have previously
identified in Figs.~\ref{fig:120_140GeV}--\ref{fig:regions_description}.

Looking at Figs.~\ref{fig:spectra}, one notices that in these regions \fermi\
model systematically (in many consecutive bins) underpredicts the data (this
is also true for the GC region).\footnote{If one excludes from fit several
  outlier bins, the resulting power law model becomes much closer to the
  predictions of \fermi background.} As \fermi diffuse model is determined
globally, one cannot improve the quality of fit in these spots without
worsening the overall fit.  In all these regions the signal is consistent with
the smooth background model plus additional sources of emission with sharp
energy spectra localized in the Galactic plane.

Indeed, let us compare two most prominent regions (\centerreg\ and \second)
(Figs.~\ref{fig:reg2} and \ref{fig:center}).  We can interpret their spectra
as a combination of spatially smooth (\fermi) background model plus a
localized excess modeled in each region by two components: a low-energy
(20-80~GeV) power law and a feature at energies around $\sim 70-80$~GeV for
\second\ and $110-130$~GeV for \centerreg regions The count rates observed in
these features (number of counts in excess of the background model) are
approximately the same, the significance of the feature is thus determined
only by the \emph{predicted background value}. This value of the predicted
signal decreases with energy, therefore the most energetic (130~GeV) feature,
having the count rate close to that of features at other energies, would have
a higher significance against very low background at that high energy.

\begin{figure*}[t]
  \centering
  \includegraphics[width=.5\textwidth]{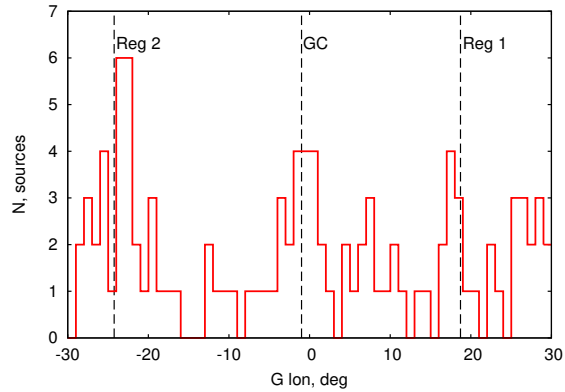}
  \caption{Number of 2FGL sources at different positions along the Galactic
    plane ($|b| \le 3^\circ$). Positions of regions \centerreg, \second\ and
    \magenta\ are clearly visible as local maxima in the number of sources.}
  \label{fig:source}
\end{figure*}

An additional evidence in favor of the ``local source'' hypothesis is provided
by the histogram of the number of point sources from the 2 year \fermi catalog
along the Galactic Plane (shown in Fig.~\ref{fig:source}).  One sees that the
regions \centerreg, \second\ and \magenta\ correspond to the local maxima in the
concentration of point sources in the Galactic plane.  This picture is
consistent with having some additional localized sources of emission in some
regions along the Galactic plane.

\bigskip

\emph{In conclusion,} at the moment there are two possible interpretations of
the observed features. In one of them the 130~GeV feature is assumed to be a
unique statistically significant feature (probably explained by DM
annihilation/decay) see e.g. \cite{weniger12,tempel12}, while all other
features are treated as fluctuations of the powerlaw background.  The other
possible interpretation is that all the features at 80~GeV, 110~GeV, 130~GeV
observed in the spectra of different parts of the Galactic plane have a
similar origin.  Our analysis shows that the present data is consistent with
``many features -- many sources'' interpretation.  If confirmed, such
interpretation would be hard to explain in terms of Dark Matter annihilation
(or decay), it would probably require an astrophysical explanation, like, for
example the one of~\cite{Aharonian:2012cs}, in which narrow spectral features
(located at different energies for different systems) are produced by emission
from ultra-relativistic pulsar wind.

If this interpretation will be confirmed by the future data, this means that a
line-like feature in gamma ray spectrum should not necessarily be ``a smoking
gun'' of dark matter annihilation (or decay), as it is often assumed.
In
Refs.~\cite{weniger12,Su-GC,Finkbeiner:12a} a procedure of data analysis is
described in which only the most significant (located in the GC) feature
survives while all other regions have their significances below
$3\sigma$. However, to speak seriously about the DM origin of a signal, one
has to show that no other interpretation is possible (for example, that the
other features may \emph{not} have physical origin). We believe that for the
moment the data does not allow to distinguish reliably between these two
interpretations. Additional observations with with HESS-2, Gamma-400 and CTA
will probably be required in order to check these models
(see~\cite{Bergstrom:2012vd}), but until that the DM interpretation of 130~GeV
feature and treating all other features in a different way looks dubious.

\subsection*{Acknowledgments}

We acknowledge useful discussions with G. Bertone, C. Frenk, G. Servant,
M.~Su. We are especially thankful to D.~Finkbeiner and C.~Weniger for fruitful
discussions, comments on our manuscript and for sharing with us their results
prior to publication. O.R. would like to thank the organizers of the IDM 2012
conference for setting up a special discussion session. The work of D.M. is
supported in part from the SCOPES project IZ73Z0\_128040 of Swiss National
Science Foundation, grant No CM-203-2012 for young scientists of National
Academy of Sciences of Ukraine, Cosmomicrophysics programme of the National
Academy of Sciences of Ukraine and by the State Programme of Implementation of
Grid Technology in Ukraine. D.M. thank the participants of ISSI team ``Present
and Past Activity of the Galactic Center Super-massive Black Hole'' for useful
discussions and the International Space Science Institute (ISSI, Bern) for
support. The authors also wish to acknowledge the SFI/HEA Irish Centre for
High-End Computing (ICHEC) for the provision of computational facilities and
support.

\def\apj{ApJ} 

\bibliography{biblio}

\end{document}